\documentclass[12pt]{article}
\usepackage{graphicx}
 \oddsidemargin  = - 7 mm 
 \evensidemargin = - 7 mm
\input{epsf.tex}

\def\jtstrut{\vrule height4ex depth0pt width0pt} 

\begin{document}
\date{}
\title{$\eta$ bound states in nuclei} 
\author{
 {\large C. Garc\'{\i}a--Recio\thanks{
e-mail: g\_recio@ugr.es}, 
J. Nieves} \\
 {\small Departamento de F\'{\i}sica Moderna, 
  Universidad de Granada, 18071 Granada, Spain}\\
 {\large T. Inoue and E. Oset} \\
 {\small Departamento de F\'{\i}sica Te\'orica and IFIC, 
  Centro Mixto Universidad de Valencia-CSIC }\\
 {\small Institutos de Investigaci\'on de Paterna,
  Apdo. correos 22085, 46071, Valencia, Spain }
       } 
\maketitle

\begin{abstract} 
 The energies and widths of bound states of the $\eta$ meson in
 different nuclei are obtained using the results for its selfenergy
 in a nuclear medium, which is evaluated in a selfconsistent manner
 using techniques of unitarized chiral perturbation theory.  We find
 bound states in all studied nuclei (from $^{12}\mbox{C}$ on) and the half
 widths obtained are larger than the separation of the levels, what
 makes the experimental observation of peaks unlikely.  We have paid
 a special attention to the region of nuclei where only the $1s$
 state appears and the binding energies are of the order of magnitude
 of the half width, which would magnify the chances that some broad
 peak could be observed.  This is found in the region of
 $^{24}\mbox{Mg}$ with a binding energy around 12.6 MeV and half
 width of 16.7 MeV.  In heavy nuclei like $^{208}\mbox{Pb}$ there are
 many bound states which would be difficult to disentangle and the
 deepest state has a binding energy about 21 MeV and half width
 around 16 MeV. Such an overlapping accumulation of states could be
 seen as an extension of the continuum of $\eta$ strength into the
 bound region in $\eta$ production experiments.

\end{abstract}

\section{Introduction}

 The $\eta$ nucleus optical potential has been a subject of intense
 study for a long time linked to the existence of $\eta$ bound states
 in nuclei~\cite{Haider:sa}-\cite{Tsushima:2000cp}.  This latter topic
 has been the object of research lately and several reactions,
 $^7\mbox{Li}(\mbox{d},^3\mbox{He})^6_{\eta}\mbox{He}$,
 $^{12}\mbox{C}(\mbox{d},^3\mbox{He})^{11}_{\eta}\mbox{B}$ and
 $^{27}\mbox{Al}(\mbox{d},^3\mbox{He})^{26}_{\eta}\mbox{Mg}$ are being
 investigated at GSI~\cite{gillitzer} and others are being carried out
 or proposed in other laboratories~\cite{Sokol:2000fv}.  With the
 exception of the $\eta$ in the two nucleon system where two
 independent works seem to rule out a bound state
 \cite{Fix:2000hf,Garcilazo:2000yy}, all the other works predict bound
 $\eta$ states from $^4$He up.  Experimentally there is no evidence
 that bound $\eta$ states exist and their observation might be
 problematic since, even if there are bound states, their widths could
 be very large compared to the separation of the levels.  This was
 indeed the case in the potential derived in Ref.~\cite{Chiang:ft}.
 However, as shown in this latter reference, there were large
 uncertainties in the potential, and hence in the energy and width of
 the bound states, mostly tied to the uncertainty in the binding
 potential of the $N^*(1535)$ resonance in the nucleus.  This has been
 an open problem for years and only recently was it possible to give a
 credible answer to this question.  The answer has come from the study
 of Refs.~\cite{nieves,inoue} where the $\pi N$ and coupled channels
 scattering was investigated up to energies above the $N^*(1535)$
 resonance, including the $\eta N$ elastic scattering of relevance for
 the $\eta$ interaction in nuclei. The study was done using techniques
 of unitarized chiral perturbation theory adapted to the meson baryon
 interaction~\cite{ramos,oller} which followed and extended the works
 using chiral Lagrangians and the Lippmann-Schwinger equation initiated
 in Refs.~\cite{wolfram,kaiser}.  In such schemes the $N^*(1535)$ resonance
 is generated dynamically from the multiple scattering of the mesons
 implied in the Lippmann-Schwinger equation of Refs.~\cite{wolfram,kaiser},
 the Bethe-Salpeter equation of Refs.~\cite{nieves,ramos} or in the N/D
 unitarization method used in Ref.~\cite{oller}. This is important, since
 by implementing systematically medium corrections to the scattering
 equation, one can see how the pole of the $N^*(1535)$ resonance is
 changed in the nuclear medium and thus eliminate the main source of
 uncertainty of earlier evaluations of the $\eta$ nucleus optical
 potential.  This latter work was conducted in Ref.~\cite{Inoue:2002xw},
 using the vacuum amplitude of Ref.~\cite{inoue}, and the $\eta$ nucleus
 optical potential was evaluated in a selfconsistent way, as it was 
 done for the antikaon case in Ref.~\cite{ro}.  Among other results, in
 Ref.~\cite{Inoue:2002xw}, it was shown that the $N^*(1535)$ resonance
 barely moved in the nuclear medium with respect to the free-space,
 although its width was certainly changed.  Besides in that reference,
 the $\eta$ selfenergy in the nuclear medium was evaluated as a
 function of the $\eta$ energy, its momentum and the nuclear density.
 In this work, we use the optical potential developed there in order 
 to calculate $\eta-$nucleus bound states, with the aim of
 finding the optimal nuclei to search for such states.  This condition
 is met when there is a minimum overlap between the states, and the
 half widths are not much bigger than the separation of the levels.  We
 find this situation for medium size nuclei rather than in the light 
 ones, which have been mostly used or suggested up to now in the search for
 such states.
 
 \section{The $\eta$ nucleus optical potential}
 
 In Ref.~\cite{Inoue:2002xw} the self-energy of the $\eta$ meson is
 evaluated in nuclear matter at various densities $\rho$, as a
 function of the $\eta$ energy, $k^0$, and its momentum, $\vec k$, in
 the nuclear matter frame.  It is calculated by means of
\begin{equation}
     \Pi_{\eta}(k^0, \vec k~; \rho) 
     =   
     4 \! \int^{k_F} \! \!  \frac{d^3 \vec p_n }{(2\pi)^3}~ 
     T_{\eta n}(P^0, \vec P ~; \rho) 
 \label{eqn:selfint}
\end{equation}
 where $\vec p_n$ and $k_F$ are the momentum of the neutron 
 and the Fermi momentum at nuclear density $\rho$ respectively, 
 and  $T_{\eta n}(P^0, \vec P ~; \rho)$ is 
 the $\eta$-neutron in-medium $s-$wave interaction,
 with the total four-momentum of the system $(P^0,\vec P)$ in the 
 nuclear matter frame, 
 namely $P^0=k^0+E_n(\vec p_n)$ and $\vec P=\vec k + \vec p_n$.
 Here, the isospin symmetry, $T_{\eta p} = T_{\eta n}$, is assumed and
 the amplitude is summed over nucleons in the Fermi sea. 
 Since we are interested in finding bound states, 
 we shall be concerned about the $s-$wave $\eta$ self-energy which around the
 $N^*(1535)$ region is largely the most relevant, given the large $s-$wave
 coupling of this resonance to the $\eta N$ states.
 
 The in-medium interaction $T_{\eta n}$ is obtained 
 by basing on the model of Ref.~\cite{inoue} for $\pi N$ and coupled channels
 scattering.
 That model reproduces the experimental data of $\pi N$ scattering
 up to energies above the $N^*(1535)$ region. 
 A similar chiral approach which covers a much wider energy range
 in isospin 1/2, although with more free parameters, 
 is also done in Ref.~\cite{nieves}.

 In Ref.~\cite{inoue}, the Bethe-Salpeter equation is considered 
 with eight coupled channels including two $\pi\pi N$ states, namely
 \{$\pi^- p$, $\pi^0 n$, $\eta n$, $K^0 \Lambda$,
   $K^+ \Sigma^-$, $K^0 \Sigma^0$,
   $\pi^0 \pi^- p$, $\pi^+ \pi^- n$\}.
 The kernels for the meson-baryon two-body sector
 are taken from the lowest order chiral Lagrangians
 and improved by applying a form factor corresponding to a vector meson
 exchange in the $t$-channel.
 The kernels for $\pi N \leftrightarrow \pi\pi N$ transitions 
 are determined so that they account for 
 both the $\pi N$ elastic and $\pi N \to \pi\pi N$ processes.
 That model reproduces well the the $\pi N$ scattering amplitudes,
 especially in isospin 1/2, for the center of mass energy energies
 from threshold to 1600 MeV.  
 It reproduces also the $\pi^- p \to \eta n$ cross section
 at the region where the $p$-wave contribution is negligible.
 In this coupled channels approach,  
 the model also provides the $\eta N$ interaction in free space
 and generates dynamically the $N^*(1535)$ resonance providing
 the width and branching ratios for its decay in good agreement with
 experiment, among them the $\eta N$ branching ratio which is
 quite large for that resonance.
 The agreement of the model with the different available data
 around the $N^*(1535)$ resonance region and the adequate description 
 of the properties of the resonance, in particular the strong coupling
 to the $\eta N$ state, give us confidence that the model is rather
 accurate to make predictions on the $\eta N \to \eta N$ interaction and $\eta$
 nucleus interaction.

 The medium effects in the $\eta N$ scattering amplitude which are
 considered in Ref.~\cite{Inoue:2002xw} are: 1) The Pauli blocking of the
 intermediate nucleon states appearing in the Bethe--Salpeter
 equation. 2) The selfenergy of the mesons (pions, kaons and eta) in
 the intermediate states, with the $\eta$ selfenergy considered
 selfconsistently. 3) The baryon selfenergy of the intermediate states
 (N, $\Lambda$ and $\Sigma$).
 
 \begin{figure}[t]
 \centerline{ 
 \epsfysize = 85 mm  \epsfbox{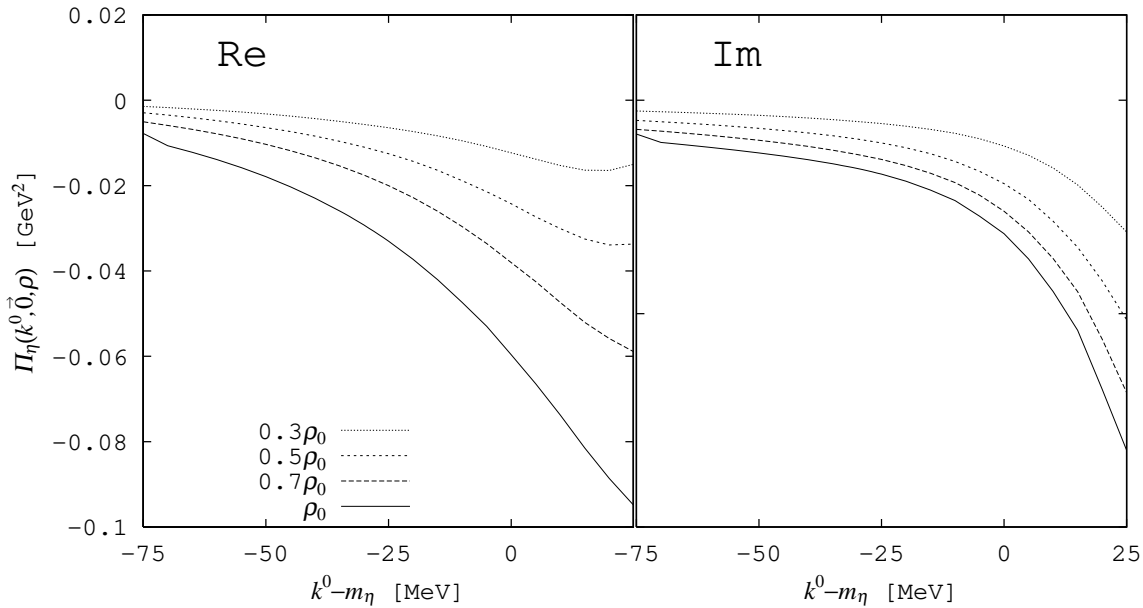}
              }
 \caption{
  $\eta$ self-energy for zero momentum as a function of energy,
  for four different densities.
  Both the Pauli blocking and hadron dressing are taken into account.
         }
 \label{fig:iiabc}
\end{figure}

 The results obtained for the $\eta$ selfenergy can be seen
 in Fig.~\ref{fig:iiabc}. There one can see that the $\eta$ selfenergy
 around zero $\eta$ energy and negative values is strongly energy dependent, 
 both for the real and the imaginary parts. 
 In order to have a feeling of the strength of the potential we can
 evaluate the optical potential by means of  
\begin{equation}
  U_{\eta}(\rho) = \frac{ \Pi_{\eta}(m_{\eta},\vec{0},\rho) }{2  m_{\eta}} ~.
\end{equation}
 This simple prescription gives the potential of $(-54 -i29)$ MeV 
 at normal nuclear matter density. 
 This means that one can expect bound states with 
 around 50 MeV binding and a width of about 60 MeV. However, the energy
 dependence of the selfenergy is quite strong as one can see from 
 Fig.~\ref{fig:iiabc} and, hence, a realistic determination of the $\eta$
 bound states should take that into account. For such purpose the results of
 Ref.~\cite{Inoue:2002xw} were parametrized in terms of analytical functions
 in the energy range $-50 ~\mbox{MeV} < k^0 - m_{\eta} < 0$ as 
\begin{eqnarray}  
 \mbox{Re}[ \Pi_{\eta}(k^0, \vec 0~; \rho) ] &=&   
       a(\rho) + b(\rho)(k^0-m_{\eta}) 
     + c(\rho)(k^0-m_{\eta})^2+ d(\rho)(k^0-m_{\eta})^3
\nonumber
\\
 \mbox{Im}[ \Pi_{\eta}(k^0, \vec 0~; \rho) ] &=&   
       e(\rho) + f(\rho)(k^0-m_{\eta}) 
     + g(\rho)(k^0-m_{\eta})^2+ h(\rho)(k^0-m_{\eta})^3
\label{eq:self}
\end{eqnarray}
 with 
\begin{eqnarray}
 a(\rho)&=& (-36200.3   ~\rho/\rho_0 -24166.6   ~\rho^2/\rho_0^2) ~\mbox{MeV}^2
\nonumber
 \\
 b(\rho)&=& (-1060.05   ~\rho/\rho_0 -326.803   ~\rho^2/\rho_0^2) ~\mbox{MeV}
\nonumber
 \\
 c(\rho)&=& -13.2403   ~\rho/\rho_0 -0.154177  ~\rho^2/\rho_0^2
\nonumber
 \\
 d(\rho)&=& (-0.0701901 ~\rho/\rho_0 +0.0173533 ~\rho^2/\rho_0^2) ~\mbox{MeV}^{-1}
\nonumber
 \\
 e(\rho)&=& (-43620.9   ~\rho/\rho_0 +11408.4   ~\rho^2/\rho_0^2) ~\mbox{MeV}^{2}
\nonumber
 \\
 f(\rho)&=& (-1441.14   ~\rho/\rho_0 +511.247   ~\rho^2/\rho_0^2) ~\mbox{MeV}
\nonumber
 \\
 g(\rho)&=& -27.6865   ~\rho/\rho_0 +10.0433   ~\rho^2/\rho_0^2 
\nonumber
 \\
 h(\rho)&=& (-0.221282  ~\rho/\rho_0 +0.0840531 ~\rho^2/\rho_0^2)
 ~\mbox{MeV}^{-1}
\label{eq:coef}
 ~~~.
\end{eqnarray}

 This potential is evaluated in infinite nuclear matter. In finite
 nuclei we use the local density approximation, substituting $\rho$ by
 $\rho (r)$, the local density at each point in the nucleus which we
 take from experiment. For the $s-$wave that we use here, it was shown in
 Ref.~\cite{Nieves:ev} that the local density approximation (LDA) gave the
 same results as a direct finite nucleus calculation.

 In the next section we solve the Klein--Gordon equation with the two
 potentials: 
 $i)$ the energy dependent one, defined in Eqs.~(\ref{eq:self}) 
 and (\ref{eq:coef}), 
 and $ii)$ an energy independent potential obtained from the latter
 one by taking $k^0 = m_\eta$ in Eq.~(\ref{eq:self}).  
 Finally, we discuss the implications of our results in the practical
 search for these $\eta$ bound states.
 
 \section{Results}

 To compute de $\eta-$nucleus bound states, we solve the Klein-Gordon 
 equation (KGE) with the strong LDA $\eta-$selfenergy, 
 $\Pi_\eta(k^0,r) \equiv \Pi_\eta(k^0,\vec 0,\rho(r))$.
 We have then:
 \begin{equation}
   \left[ -\vec\nabla^2 + \mu^2 +  \Pi_\eta(\mbox{Re}[E],r) \right] \Psi 
  = E^2 \Psi
 \end{equation}
 where $\mu$ is the $\eta-$nucleus reduced mass, the real part of $E$  
 is the total meson energy, including its mass, and the imaginary part
 of $E$, with opposite sign, is the half-width $\Gamma/2$ of the state. 
 The binding energy $B<0$ is defined as $B = \mbox{Re}[E]-m_{\eta}$. 
 As mentioned above two different $\eta-$selfenergies are being considered.
 
 In order to solve the KGE in coordinate space we use a numerical algorithm
 which has been extensively tested in the similar problems of
 pionic~\cite{Nieves:ev} and antikaonic~\cite{baca} atomic states and in
 the search of possible antikaon-nucleus bound states~\cite{baca}.
 Charge densities are taken from Ref.~\cite{Ja74}.
 For each nucleus, we take the neutron matter
 density approximately equal to the charge one,
 though we consider small changes, inspired by Hartree-Fock
 calculations with the DME (density-matrix expansion)~\cite{Ne75} and
 corroborated by pionic atom analysis~\cite{nog92}. 
 In Table~1 of Ref.~\cite{baca} all the densities used throughout this
 work can be found.  
 However, charge (neutron matter) densities do not correspond
 to proton (neutron) ones because of the finite size of the proton (neutron). 
 We take that into account following the lines of Ref.~\cite{Nieves:ev}
 and use the proton (neutron) densities in our approach.
  
\begin{figure}[p]
\vspace{-1.cm}
\begin{center}                                                                
\leavevmode
\makebox[0cm]{
\epsfysize = 300pt
\epsfbox{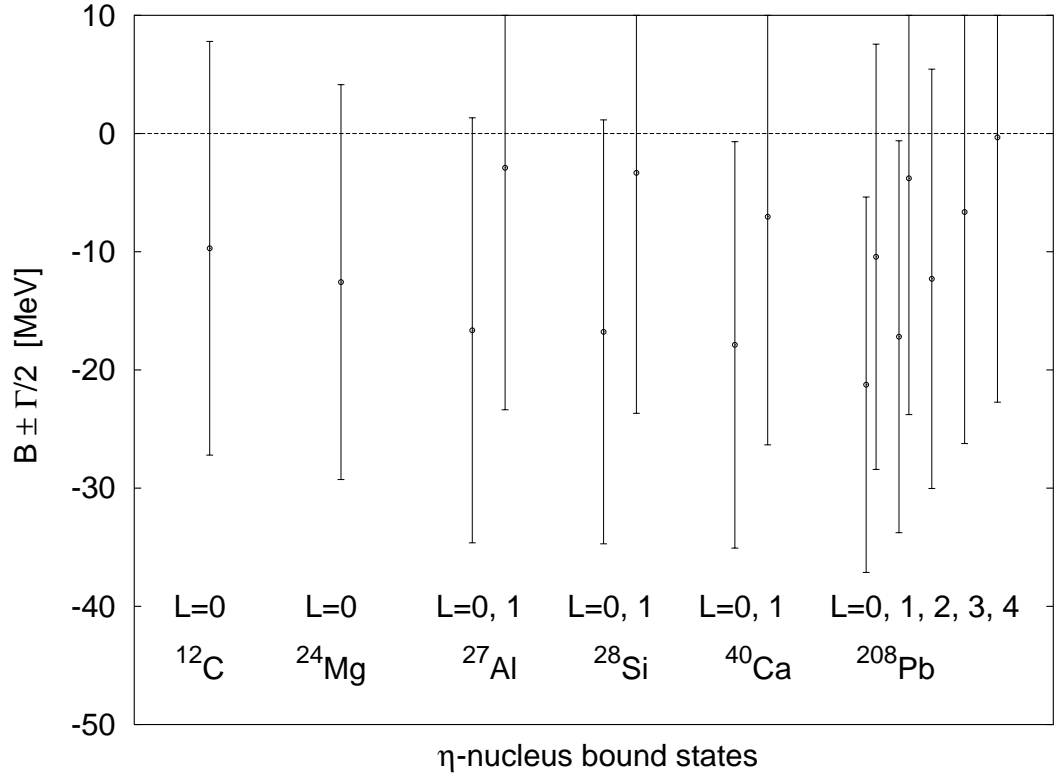}}\\ 
\vspace*{.1cm} 
\makebox[0cm]{
\epsfysize = 300pt
\epsfbox{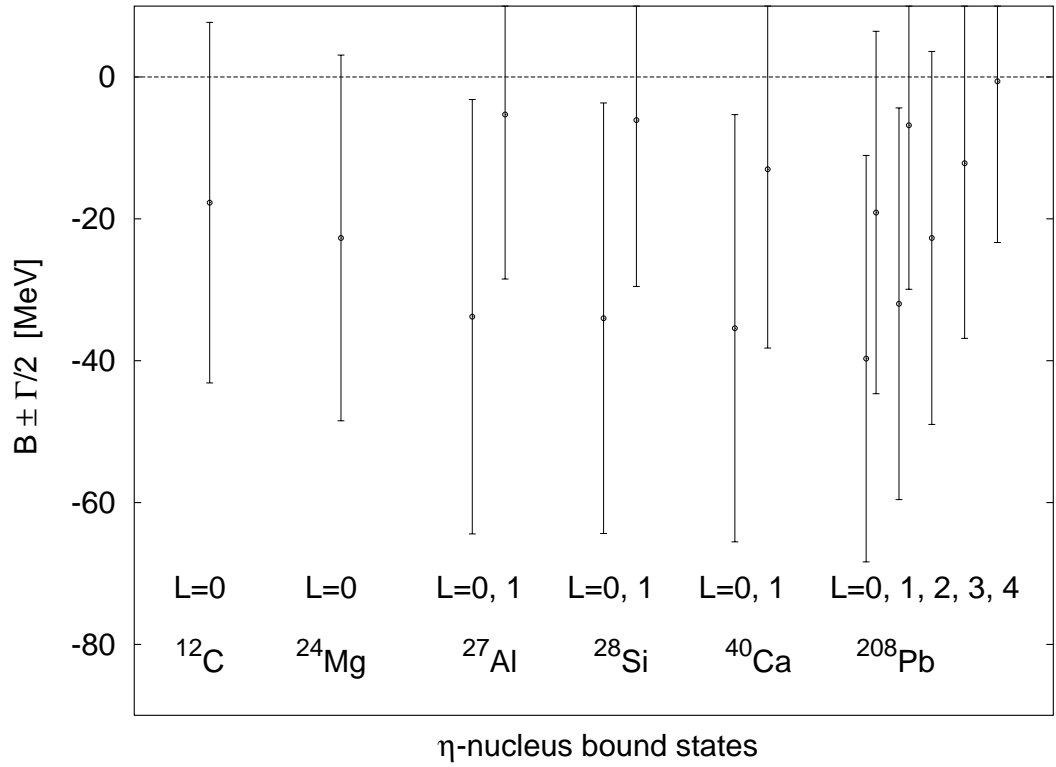}}
\end{center}
\caption {\small Top panel: Binding energies and widths for different
  nuclei obtained using the strong energy dependent $\eta-$selfenergy
  $i)$. Bottom panel: Results obtained with the energy independent
  $\eta-$selfenergy $ii)$ .}
\label{fig:i}
\end{figure}

 Our results are shown in Tables 1 and 2 for the energy dependent and energy
 independent potentials respectively. We also show them in graphic form 
 in Fig.~\ref{fig:i}. As an average we can see that the binding energies are
 about double with the energy independent potential than with the energy
 dependent one, which shows the importance
 of taking into account this effect. On the
 other hand we see that the half widths of the states are  
 large, larger in fact
 than the binding energies or the separation energies between neighboring 
 states. The widths are also smaller with the energy dependent potential, but
 comparatively  to the size of the binding energies they are 
 a little bit larger.
 
\begin{table}[t]
\begin{center}
\caption{ (B,$-\Gamma/2$) for $\eta-$nucleus bound states 
calculated with the energy dependent potential.}
\label{tbl:statedep}
\footnotesize
\begin{tabular}{c|cccccc}
\hline
\hline \jtstrut
  &   $^{12}$C  & $^{24}$Mg    &  $^{27}$Al    &  $^{28}$Si    &   $^{40}$Ca   &  $^{208}$Pb    \\
\hline \jtstrut
1s&($-$9.71,$-$17.5)&($-$12.57,$-$16.7)&($-$16.65,$-$17.98)&($-$16.78,$-$17.93)&($-$17.88,$-$17.19)&($-$21.25,$-$15.88) \\
1p&             &              &( $-$2.90,$-$20.47)&( $-$3.32,$-$20.35)&( $-$7.04,$-$19.30)&($-$17.19,$-$16.58) \\
1d&             &              &               &               &               &($-$12.29,$-$17.74) \\
2s&             &              &               &               &
&($-$10.43,$-$17.99) \\
1f&             &              &               &               &               &( $-$6.64,$-$19.59) \\
2p&             &              &               &               &               &( $-$3.79,$-$19.99) \\
1g&             &              &               &               &               &( $-$0.33,$-$22.45) \\
\hline
\hline
\end{tabular}
\normalsize
\end{center}
\end{table}

\begin{table}[t]
\begin{center}
\caption{(B,$-\Gamma/2$) for $\eta-$nucleus bound 
states calculated with the energy independent potential.}
\label{tbl:stateind}
\footnotesize
\begin{tabular}{c|cccccc}
\hline
\hline
\jtstrut
  &   $^{12}$C    & $^{24}$Mg     &  $^{27}$Al     &  $^{28}$Si     &
$^{40}$Ca   &  $^{208}$Pb   \\ 
\hline \jtstrut
1s&($-$17.71,$-$25.42)&($-$22.69,$-$25.78)&($-$33.80,$-$30.63)&($-$34.01,$-$30.36)&($-$35.42,$-$30.12)&($-$39.71,$-$28.65) \\
1p&               &               &( $-$5.28,$-$23.20)&(
$-$6.07,$-$23.45)&($-$13.02,$-$25.19)&($-$31.97,$-$27.61) \\
1d&               &               &                &                &             &($-$22.69,$-$26.30) \\
2s&               &               &                &                &             &($-$19.11,$-$25.55) \\
1f&               &               &                &                &             &($-$12.16,$-$24.69) \\
2p&               &               &                &                &             &( $-$6.81,$-$23.12) \\
1g&               &               &                &                &             &( $-$0.60,$-$22.74) \\
\hline
\hline
\end{tabular}
\normalsize
\end{center}
\end{table}

   With the results obtained here it looks like the chances to see distinct
 peaks corresponding to $\eta$ bound states are not too big.  However, the
 systematics carried out with the study of different nuclei is instructive. 
 Indeed,
 we see that the tendency for nuclei lighter than $^{24}\mbox{Mg}$ is that the 
 binding
 energy becomes smaller but the half width stays comparatively more stable,
 such that the ratio of the width to binding energy becomes bigger for lighter
 nuclei and hence the chances to see these bound states become smaller. 
 This tendency seems to be telling us that the light nuclei is
 definitively not the place where one should search for $\eta$ bound states.
 On the other hand,  if we go to heavy nuclei, 
 like $\mbox{Pb}$, there is a superposition of many bound states and the
 separation of the levels is small compared to the half width of these states.
 The chances to see peaks there are null. On the other hand there is a region,
 which, even with difficulty, 
 still provides the optimum ground to see the $\eta$ bound states.
 This appears around the $\mbox{Mg}$ region where the binding energy is
 similar to the half width and there is only one bound state. If one goes to
 heavier nuclei the ratio of binding energy versus width become bigger,
 a welcome feature for the observation of the bound states, 
 but simultaneously there appear new bound states, 
 as one can see in $^{27}\mbox{Al}$, such that the half width of the states
 is much bigger than the separation between the levels.
 With all these considerations one comes to the conclusion
 that the region of nuclei around $^{24}\mbox{Mg}$ would offer
 the best chances to see the $\eta$ bound states. 
 In this case, with a binding energy of around 12.5 MeV and
 a half width of 16.7 MeV one could see still some broad bump.
 
   On the other hand one can look at the results with a more optimistic view
 if one simply takes into account that experiments searching for these states
 might not see them as peaks, but they should see some clear
 strength below threshold in the $\eta$ production experiments.
 The range by which this strength would go into the bound region
 would measure the combination of half width and binding energy. 
 Even if this is less information than the values of the energy and
 width of the states, it is by all means a relevant information to gain some
 knowledge on the $\eta$ nucleus optical potential.
 
\section{Conclusions}

 We have used a recent $\eta$ nucleus optical potential, evaluated within a 
 unitarized chiral perturbative approach, in order to find $\eta$ bound states.
 The potential is attractive and produces bound states for all nuclei which
 we study from $^{12}\mbox{C}$ up. 
 On the other hand it also produces large widths, with the half widths
 slightly larger than the binding energy, which makes the observation
 of these states unlikely. We calculated the results using an energy
 independent potential, which is obtained taking the value of the potential for
 an $\eta$ energy equal to its mass. 
 The second, more realistic potential, takes into account 
 the strong energy dependence of the potential due to the fact that
 it is much influenced by the $N^*(1535)$ resonance appearing above the 
 $\eta N$ threshold. This potential reduces the strength of both the real and
 imaginary parts of the potential for energies below the $\eta$ threshold
 and leads to substantially narrower states, 
 but at the same time with smaller binding energies, 
 such that the half widths of the states are still larger than the
 binding energies. We found that light nuclei are not the best ones to search
 for $\eta$ bound states since the widths become comparatively much larger
 than the binding energies. 
 The heavy nuclei accommodate many bound states and the separation
 of the levels is much smaller than the half width of the states. 
 The best chances for observation of bound states are 
 in the region of $^{24}\mbox{Mg}$ where there is only one bound state
 and the half width is only a little bigger than the binding energy.
 In any case it was stressed that, even if no broad bumps are found
 in the experiments, they should find some strength in the bound region 
 stretching in energy down to the sum of the binding energy plus the half 
 width of the bound states. Short of having the values for the binding energy 
 and width of the states, this more limited information is still very 
 valuable to gain some knowledge on the $\eta$ nucleus optical potential
 and it should stimulate experiments in this direction.

\section*{Acknowledgments}
 
 This work has been partly supported by the Spanish Ministry of Education
 in the program 
 ``Estancias de Doctores y Tecn\'ologos Extranjeros en Espa\~na'',
 by the DGICYT contract number BFM2000-1326,  the DGES contract
 number PB98-1367, the EU TMR network Eurodaphne, contact
 no. ERBFMRX-CT98-0169, and by the Junta de Andaluc\'{\i}a.

\end{document}